\documentclass[11pt]{article}
\usepackage{amsfonts}
\usepackage{amsmath,amssymb}

\newcommand{\hko}{\hookrightarrow}

\newcommand{\n}{\nonumber\\}

\newcommand{\bec}{\begin{center}}
\newcommand{\eec}{\end{center}}

\newcommand{\bea}{\begin{array}}
\newcommand{\ear}{\end{array}}

\newcommand{\bfr}{\begin{flushright}}

\newcommand{\efr}{\end{flushright}}
\newcommand{\noi}{\noindent}
\newcommand{\me}{\frac{1}{2}}

\newcommand{\cl}{{\mt{C}}\ell}

\newcommand{\RR}{\mathbb{R}}\newcommand{\op}{\oplus}
\newcommand{\HH}{\mathbb{H}}
\newcommand{\ap}{\alpha}
\newcommand{\be}{\beta}

\newcommand{\la}{\Lambda}
\newcommand{\bege}{\begin{equation}}
\newcommand{\enge}{\end{equation}}
\newcommand{\w}{\wedge}
\newcommand{\g}{\gamma}
\newcommand{\om}{\omega}
\newcommand{\ri}{\rightarrow}

\newcommand{\si}{\sigma}

\newcommand{\beq}{\begin{eqnarray}}\newcommand{\benu}{\begin{enumerate}}\newcommand{\enu}{\end{enumerate}}
\newcommand{\eeq}{\end{eqnarray}}
\newcommand{\mt}{\mathcal}

\newcommand{\vv}{{\bf v}}
\newcommand{\ee}{{\bf e}}
\newcommand{\uu}{{\bf u}}

\newcommand{\pa}{\partial}

\newcommand{\CC}{\mathbb{C}}

\newcommand{\stk}[1]{\stackrel{#1}{\longrightarrow}}

\newcommand{\ul}{\underline}
\newcommand{\ol}{\overline}
\newcommand{\vbn}{\blacktriangleleft}
\newcommand{\vvn}{\blacktriangleright}\newcommand{\Ll}{\longleftrightarrow}

\newcommand{\KKKK}{{\bf{\mathcal{K}}}}\newcommand{\LL}{{\bf{\mathcal{L}}}}

\newcommand{\ip}{\mathfrak{I}}

\newcommand{\clt}{{\mt{C}}\ell_{3,0}}
\newcommand{\cle}{{\mt{C}}\ell_{1,3}}

\newcommand{\bx}{\begin{pmatrix}}
\newcommand{\ex}{\end{pmatrix}}

\newcommand{\mma}{{\mathfrak{a}}}
\newcommand{\mmb}{{\mathfrak{b}}}

\newcommand{\mme}{{\mathfrak{e}}}

\newcommand{\GG}{\mathbb{G}}
\newcommand{\mmf}{{\mathfrak{f}}}

\usepackage{amsfonts}
\usepackage{amsmath,amssymb}

\usepackage{indentfirst}  
\begin{document}
\title{Revisiting Clifford algebras and spinors II:\\ Weyl spinors in $\cl_{3,0}$ and $\cl_{0,3}$ and the Dirac equation}
\author{{\bf Rold\~ao da Rocha}\thanks{Instituto de F\'{\i}sica Gleb Wataghin (IFGW), Unicamp, CP 6165, 13083-970, Campinas (SP), Brazil. 
E-mail: roldao@ifi.unicamp.br. Supported by CAPES.}\and{\bf Jayme Vaz, Jr.}\thanks{Departamento de Matem\'atica Aplicada, IMECC, Unicamp, CP 6065, 13083-859, Campinas (SP), Brazil. E-mail: 
vaz@ime.unicamp.br}}
\date{}\maketitle

\abstract${}^{}$\begin{center}
\begin{minipage}{12cm}$ \;\;\;\;\;$
This paper is the second one of a series of three and it is the continuation 
of \cite{rol1}. We review some properties of the algebraic spinors in
$\clt$ and $\cl_{0,3}$ and how  Weyl, Pauli and Dirac spinors are constructed in $\cl_{3,0}$ (and $\cl_{0,3}$, in the case of Weyl spinors). 
A plane wave solution
for the Dirac equation is obtained, and the  Dirac equation is written in terms of Weyl spinors, and alternatively, in terms of Pauli spinors.
Finally  the covariant and contravariant undotted spinors in $\cl_{0,3}\simeq\HH\op\HH$ are constructed. We prove that there exists 
 an application that maps $\cl_{0,3}^+$, viewed as a right $\HH$-module, onto $\cl_{0,3}^+$,
but now viewed as a left $\HH$-module.

\end{minipage}\end{center}
\medbreak
\medbreak\noi\newpage
\section*{Introduction}

The theory of spinors was developed practically in an independent way by mathematicians and physicists. 
On the one hand, E. Cartan in 1913 wrote a treaty about spinors \cite{cartan}, after he has originally discovered them as entities that
carry representations of the rotation groups associated to finite-dimensional vector spaces.
 He was  investigating linear representations of simple groups. On the other hand, spinors were introduced in physics
in order to describe the wave function of quantum systems with spin. W. Pauli, in 1926,  described the electron wave function with spin
by a 2-component spinor in his non-relativistic theory \cite{pauli}. After, in 1928,  P. A. M. Dirac used a 4-component spinor to investigate 
 the relativistic quantum mechanics formalism \cite{dirac}. 
With the increasing use of spinors in physical theories,  L. Infeld and B. L van der Waerden \cite{inf} wrote a treaty,  
but their formalism is not so simple for an undergraduate student in mathematics, or physics,  to learn. 
In physical theories, spinors are fundamental entities describing matter, constituted by leptons and quarks, since they are spin 1/2 fermions \cite{huang}
naturally described by {\it Dirac spinors}\footnote{Dirac spinors are defined as the sum of two elements, called \emph{Weyl spinors},
 that respectively  carry two non-equivalent
representations  of the group SL(2,$\CC$).}.
From the algebraic viewpoint, spinors are elements of a lateral minimal ideal of a Clifford algebra. This was introduced by C. Chevalley \cite{chev}.  

The main aim of this paper is to formulate the paravector model of spacetime, using algebraic Weyl spinors,
and to describe Weyl spinors in the Clifford algebras $\cl_{3,0}$ and $\cl_{0,3}$. Consequently Dirac spinors
are naturally introduced, together with  (algebraic) Pauli spinors\footnote{Pauli spinors are elements of the representation space of the group SU(2).}. 
This paper is organized as follows:
In Sec. 1, we present some brief mathematical preliminaries concerning Clifford algebras. In Sec. 2 
contravariant and covariant, dotted and undotted  Weyl spinors are introduced in $\cl_{3,0}$, 
together with the spinorial transformations associated to each one of them. In this way, Dirac spinors
are naturally presented as elements of the Clifford algebra $\clt$ over $\RR^3$. 
In Sec. 3 the Dirac-Hestenes equation (DHE) in $\clt$ is introduced.
DHE is written as two coupled Weyl equations, using Weyl spinors, and alternatively, using Pauli spinors. 
The null tetrad in spacetime is introduced, using Weyl spinors to construct the paravector model of spacetime.  
In Sec. 4 we construct contravariant and covariant (Weyl) dotted spinors in
 $\cl_{0,3}\simeq \HH\oplus\HH$ and define an application that maps $\cl_{0,3}^+$, viewed as a right module (over the quaternion ring $\HH$) onto $\cl_{0,3}^+$,
but now viewed as a left $\HH$-module.  

\section{Preliminaries}
Let $V$ be a finite $n$-dimensional real vector space. We consider the tensor algebra $\bigoplus_{i=0}^\infty T^i(V)$ from which we
restrict our attention to the space $\Lambda(V) := \bigoplus_{k=0}^n\Lambda^k(V)$ of multivectors over $V$ ($\Lambda^k(V)$
 denotes the space of the antisymmetric
 $k$-tensors).   The \emph{reversion} of $\psi\in \Lambda(V)$, denoted by $\tilde{\psi}$, is an algebra antiautomorphism
 given by $\tilde{\psi} = (-1)^{[k/2]}\psi$ ([$k$] denotes the integer part of $k$)  while the \emph{main automorphism} or \emph{graded involution}\footnote{Or parity
operator.} of $\psi$, denoted by 
 $\hat{\psi}$ , is an algebra automorphism given by 
$\hat{\psi} = (-1)^k \psi$. The \emph{conjugation}, denoted by $\underline{\psi}$, is defined to be the reversion followed by the main automorphism.
  If $V$ is endowed with a non-degenerate, symmetric, bilinear map $g: V\times V \rightarrow \RR$, it is 
possible to extend $g$ to $\la(V)$. Given $\psi=u_1\w\cdots\w u_k$ and $\phi=v_1\w\cdots\w v_l$, $u_i, v_j\in V$, one defines $g(\psi,\phi)
 = \det(g(u_i,v_j))$ if $k=l$, and $g(\psi,\phi)=0$ if $k\neq l$. Finally, the projection of a multivector $\psi= \psi_0 + \psi_1 + \cdots + \psi_n$,
 $\psi_k \in \la^k(V)$, on its $p$-vector part is given by $\langle\psi\rangle_p$ := $\psi_p$. 
The Clifford product between $v\in V$ and $\psi\in\la(V)$ is given by $v\psi = v\w \psi + v\cdot \psi$.
 The Grassmann algebra $(\la(V),g)$ 
endowed with this product is denoted by $\cl(V,g)$ or $\cl_{p,q}$, the Clifford algebra associated to $V\simeq \RR^{p,q},\; p + q = n$.

\section{Weyl spinors in $\clt$}

In this section Weyl spinors and spinorial metrics are constructed.

\subsection{Weyl spinors and spinor metrics} 

\label{subsec3.1.3}

Let $\{\ee_1, \ee_2, \ee_3\}$ be an orthonormal basis of $\RR^3$.  The Clifford algebra 
$\clt$ is generated by  $\{1, \ee_1, \ee_2, \ee_3\}$, such that 
\bege
2g(\ee_i, \ee_j) = 2\delta_{ij} = (\ee_i\ee_j + \ee_j\ee_i), \quad i,j = 1, 2, 3.
\enge
An arbitrary element of $\clt$ can be written as
\bege{\label{psi}}
\psi = s + a^1\ee_1 + a^2\ee_2 + a^3\ee_3 + b^{12}\ee_{12} + b^{13}\ee_{13} +  b^{23}\ee_{23} + p\;\ee_{123}, \quad s, a^i, b^{ij}, p \in \RR.
\enge
\noi

Let $f_\pm = \frac{1}{2}(1 \pm \ee_3)$ be primitive idempotents of $\clt$, clearly satisfying the relations
 $f_+ f_- = f_- f_+ = 0$ and $f_\pm^2 = f_\pm$. The matrix representation  
\bege\label{rep11}
f_+ = \left(\begin{array}{cc}
               					1&0\\
               					0&0
               					\end{array}\right), \quad f_- =  \left(\begin{array}{cc}
               					0&0\\
               					0&1
               					\end{array}\right),\quad
               					\ee_1 f_+ = \left(\begin{array}{cc}
               					0&0\\
               					1&0
               					\end{array}\right), \quad \ee_1f_- =  \left(\begin{array}{cc}
               					0&1\\
               					0&0
               					\end{array}\right) \enge\noi is used heretofore. Each one of the four elements above
generates a (left or right) minimal ideal.

The isomorphism  
\bege\label{isofmais}
\cl_{3,0} f_+ \simeq \cl_{3,0}^+ f_+
\enge\noi is not difficult to see, since
\beq
\cl_{3,0}^+f_+ \ni \phi_+ f_+ &=& \left(\begin{array}{cc}
               					w_1&-w_2^*\\
               					w_2&w_1^*
               					\end{array}\right)f_+\nonumber\\ &=& \left(\begin{array}{cc}
               					w_1&0\\
               					w_2&0
               					\end{array}\right) \simeq  \left(\begin{array}{cc}
               					w_1&w_3\\
               					w_2&w_4
               					\end{array}\right)\left(\begin{array}{cc}
               					1&0\\
               					0&0
               					\end{array}\right) \in \cl_{3,0}f_+.
\eeq
\noindent Then there is redundancy if $\psi f_+$, with $\psi$ given by eq.(\ref{psi}), is written.  From the isomorphism in (\ref{isofmais}) 
 we need only to use  $\psi f_+$, with
 $\psi = s + b^{12}\ee_{12} + b^{13}\ee_{13} + b^{23}\ee_{23} \in \cl_{3,0}^+$. 
We now define the Weyl spinors \cite{fig}:
 \medbreak$\bullet$ {\ul{\it Contravariant undotted spinor}} (CUS):
\bege\label{scta}\bea{|l|}\hline{\rule[-3mm]{0mm}{8mm} 
\KKKK = \psi f_+}\\ \hline\ear
\enge\noi  Such spinor is written as
\beq
 \KKKK = \psi f_+ &=& (s + b^{12}\ee_{12} + b^{13}\ee_{13} + b^{23}\ee_{23})f_+\nonumber\\
 &=& (s + b^{12}\ee_{123})(f_+) + (b^{13} + b^{23}\ee_{123})(\ee_1 f_+)\nonumber\\
  &=&  k^1(f_+) + k^2(\ee_1 f_+),\eeq  \noindent where
\beq
k^1 &=& s + b^{12}\ee_{123}\quad{\rm and}\n
k^2 &=&  b^{13} + b^{23}\ee_{123}.\eeq \noi The CUS are chosen to be written in this form because their
components commute with the basis $\{f_+, \ee_1f_+\}$ of the algebraic spinors. Therefore all spinor components 
are written as elements of the center of $\cl_{3,0}$, that is well-known to be isomorphic to $\Lambda^0(\RR^3)\oplus\Lambda^3(\RR^3)$. 

From  the spinor $\KKKK$, other three types of spinors in $\clt$ are constructed: 
\medbreak$\bullet$ {\ul{\it Covariant undotted spinor}} (CVUS): 
 
\bege\bea{|l|}\hline{\rule[-3mm]{0mm}{8mm} \KKKK^* := \ee_1 \ul{\KKKK}}\\ \hline\ear\enge 
By this definition, the expression 
 \beq \KKKK^* &=&\ee_1\ul{(k_1f_+ + k_2 \ee_1 f_+)}\n
 &=& \ee_1 (f_- k^1 + f_- (-\ee_1)k^2)\n
  &=&  (-k^2)f_+ + (k^1)(f_+\ee_1)\eeq\noi can immediately be written.
Since $\KKKK^* \in f_+\cl_{3,0}$, we write
\bege\KKKK^* = k_1(f_+) + k_2 (f_+\ee_1),\enge\noi from where the relation 
\bege\bea{|l|}\hline{\rule[-3mm]{0mm}{8mm}
k_1} = -k^2\\
k_2 = k^1\\ \hline\ear
\enge\noi follows.
\noindent Note that these relations are the ones obtained in the classical approach \cite{inf,pe2}.

Now, given $\KKKK^* \in f_+\cl_{3,0}$ and $ \eta = \eta^1 f_+ + \eta^2 \ee_1 f_+\in\cl_{3,0} f_+ $, the {\it spinorial metric}
associated to the idempotent  $f_+$ is defined: 
\beq G_{f_+}:f_+ \cl_{3,0} \times \cl_{3,0} f_+ &\rightarrow& f_+ \cl_{3,0} f_+ \simeq \CC f_+\nonumber\\ 
(\KKKK, \eta)&\mapsto&  \KKKK^* \eta =  (-k^2f_+ + k^1f_+\ee_1) (\eta^1 f_+ + \eta^2 \ee_1 f_+),
\eeq\noi which results in the expression
\bege\bea{|l|}\hline{\rule[-3mm]{0mm}{8mm} 
 G_{f_+}(\KKKK, \eta) = \KKKK^* \eta = (-k^2\eta^1 + k^1\eta^2)(f_+)}\\ \hline\ear
\enge
\noi This definition coincides with the classical one \cite{inf,pe2}, where the scalar product 
has mixed and antisymmetric  components. The idempotent  $f_+$ is the unit of the algebra $f_+\cl_{3,0}f_+ \simeq \CC$.

From the spinor $\KKKK$ we also define the 
\medbreak $\bullet$ {\ul{\it Contravariant dotted spinor}} (CDS): 
\bege\label{sctp} \bea{|l|}\hline{\rule[-3mm]{0mm}{8mm} \ol\KKKK := \ee_1 \tilde{\KKKK}}\\ \hline\ear\enge
\noi This definition  results in the expression
   \beq{\ol{\KKKK}} &=& \ee_1(k^1f_+ + k^2\ee_1 f_+)^{\widetilde{\;}}\n
 &=& \ee_1(f_+{\tilde{k^1}} + f_+ \ee_1 {\tilde{k^2}})\n
&=&{\tilde{k^1}}(\ee_1 f_+) + {\tilde{k^2}}f_-\nonumber\\ &=& {\tilde{k^1}} (f_-\ee_1) + {\tilde{k^2}}(f_-).\eeq
But  $\ol{\KKKK} \in f_-\cl_{3,0}$, and it is possible to write
\bege
\ol{\KKKK} = {\ol{k}}^{1'}(f_-\ee_1) + {\ol{k}}^{2'}(f_-).
\enge\noi Then the relation
\bege\bea{l} 
{\ol{k}}^{1'} = \tilde{k^1},\\ 
{\ol{k}}^{2'} = \tilde{k^2}\ear\noi 
\enge\noi is obtained. Besides\footnote{${\bf A} = {1, 2}$.}, ${\tilde{k^{\bf A}}} = 
(a + b\ee_{123})^{\widetilde{\;}} = (a + b\ee_{321}) = a - b\ee_{123}$, which suggests the notation\footnote{
Denoting $\Lambda^0(\RR^3)$ the subspace of the scalars and  $\Lambda^3(\RR^3)$ the space of the pseudoscalars,
the isomorphism $\Lambda^0(\RR^3) \oplus \Lambda^3(\RR^3) \simeq \CC$ is evident, since $(\ee_{123}^2 = -1)$.
The notation $\tilde{k^{\bf A}}= \ol{k^{\bf A}}$ is also immediate, since if  $\ee_{123}$ is denoted by
 ${\mathfrak{I}}\simeq i\in\CC$,
the reversion in  $\Lambda^0(\RR^3) \oplus \Lambda^3(\RR^3)$ is equivalent to the $\CC$-conjugation.} 

\bege\tilde{k^{\bf A}}= \ol{k^{\bf A}}.\enge

Finally the \medbreak$\bullet$ {\ul{\it Covariant dotted spinor}} (CVDS) is constructed:
\bege\bea{|l|}\hline{\rule[-3mm]{0mm}{8mm} 
{\ol{\KKKK}}^* := \ul{(\ee_1\ol{\KKKK})}}\\ \hline\ear\enge
\noi from where it can be shown that
\beq\ol{\KKKK}^* &=& \ul{(\ee_1\ol{\KKKK})}\n
 &=& -\ul{(\ol{\KKKK})}\ee_1\n  &=&
 -\ul{{\ol{k}}^{1'}(f_-\ee_1) + {\ol{k}}^{2'} (f_-)}\ee_1\n &=&
 -(-\ee_1 f_+{\ol{k}}^{1'} + f_+ {\ol{k}}^{2'})\ee_1 \nonumber\\&=&{\ol{k}}^{1'}f_- - {\ol{k}}^{2'}f_+\ee_1  \nonumber\\ &=&  (-{\ol{k}}^{2'})(\ee_1f_-) + ({\ol{k}}^{1'})(f_-).\eeq
As  $\ol{\KKKK}^* \in \cl_{3,0} f_-$, we can write
\bege
\ol{\KKKK}^* = (\ol{k}_{1'})(\ee_1f_-) + (\ol{k}_{2'})(f_-).
\enge\noi Therefore the following relation is obtained:
\bege\bea{|l|}\hline{\rule[-3mm]{0mm}{8mm} 
\ol{k}_{1'} = -{\ol{k}}^{2'}}\\
\ol{k}_{2'} = {\ol{k}}^{1'}\\ \hline\ear
\enge\noi These relations are the ones obtained in van der Waerden paper \cite{inf}, and unceasingly exposed in Penrose seminal works \cite{pe2,pe1,pe3,pe4,pe5}.

Now, given  $\ol{\KKKK} \in f_-\cl_{3,0}$ and
 $ \ol{\eta}^* = \ol{\eta}^{2'}\ee_1 f_+ + \ol{\eta}^{1'} f_-\in\clt f_-$, the {\it spinorial metric} associated to
the idempotent $f_-$ is obtained: 
\beq G_{f_-}:f_- \cl_{3,0} \times \cl_{3,0} f_- &\rightarrow& f_- \cl_{3,0} f_- \simeq \CC f_-\nonumber\\
({\ol{\KKKK}}, {\bar{\eta}}^*)&\mapsto& \ol{\KKKK}{\ol{\eta}}^* =  ({\ol{k}}^{1'}f_-\ee_1 + {\ol{k}}^{2'}f_-)(-{\ol{\eta}}^{2'}\ee_1 f_- + {\ol{\eta}}^{1'} f_-),
\eeq which results in 
\bege\bea{|l|}\hline{\rule[-3mm]{0mm}{8mm}
G_{f_-}=\ol{\KKKK}{\ol{\eta}}^* = (\ol{\eta}^{1'}\ol{k}^{2'} - \ol{\eta}^{2'}\ol{k}^{1'}) f_-}\\ \hline\ear
\enge
The expressions for the four algebraic Weyl spinors, as elements of a lateral ideal of $\cl_{3,0}$, are listed below:
\begin{itemize}
\item {\ul{\it Contravariant undotted spinor}} (CUS): 
\bege\bea{|l|}\hline{\rule[-3mm]{0mm}{8mm}
\KKKK = k^1(f_+) + k^2(\ee_1 f_+) \in \cl_{3,0} f_+}\\ \hline\ear\enge

\item {\ul{\it Covariant undotted spinor}} (CVUS): 
\bege\bea{|l|}\hline{\rule[-3mm]{0mm}{8mm}
\KKKK^* = \ee_1 \ul{\KKKK} = k_1(f_+) + k_2 (f_+\ee_1) \in f_+\cl_{3,0}}\\ \hline\ear\enge

\item {\it\ul{Contravariant dotted spinor}} (CDS): \bege\bea{|l|}\hline{\rule[-3mm]{0mm}{8mm}\ol\KKKK = \ee_1 \tilde{\KKKK} = {\ol{k}}^{1'}(f_-\ee_1) + {\ol{k}}^{2'} (f_-) \in f_-\cl_{3,0}}\\ \hline\ear\enge

\item {\ul{\it Covariant dotted spinor}} (CVDS): 
\bege\bea{|l|}\hline{\rule[-3mm]{0mm}{8mm}\ol{\KKKK}^* = -(\ul{\ol{\KKKK}})\ee_1 = (\ul{\ee_1\ol{\KKKK}}) = \ol{k}_{1'}(\ee_1 f_-) + {\ol k_{2'}} (f_-) \in \cl_{3,0} f_-}\\ \hline\ear
\enge
\end{itemize}
The following diagram illustrates how we can pass from one ideal to the others, obtaining in this way all the four Weyl spinors,
using (anti-)automorphisms of $\clt$:
\vspace{1cm}
$$
\begin{array}{|ccccccc|}\hline{\rule[-3mm]{0mm}{8mm}
\KKKK} & \stk{*} & {\KKKK}^* & \stk{\widehat{}} & \ol{\KKKK} & \stk{*} &  {\ol{\KKKK}}^* = \widehat{\KKKK}\\
\uparrow& &\uparrow& &\uparrow& &\uparrow\\ 
contravariant & &covariant & &contravariant & & covariant\\
undotted & &undotted& &dotted& &dotted\\
\cl_{3,0} f_+& &f_+\cl_{3,0}& &f_- \cl_{3,0}& & \cl_{3,0}f_-\\ \hline
\end{array}
$$
\vspace{1cm}

We then have the correspondence between the formalism of this section and the notation exhibited in  \cite{pe2}:

\bege\begin{array}{|l|}\hline{\rule[-3mm]{0mm}{8mm}
\GG^A \Ll\clt f_+,\quad\quad \GG_A\Ll f_+\clt,\quad\quad\GG^{A'} \Ll f_-\clt,\quad\quad\GG_{A'}\Ll\clt f_-}\\ \hline\ear\enge

\subsection{Spinorial tranformations}
An arbitrary element $R\in\cl_{3,0}$ can be written as
\bege R = s + v^i\ee_i + b^{ij}\ee_{ij} + p\ee_{123} = \ap + \be\ee_{12} + \gamma\ee_{13} + \delta\ee_{23},\enge
\noindent where $\ap = s + p\ee_{123},\;\; \be = b^{12} - v^3\ee_{123}, \;\;\g = b^{13} + v^2\ee_{123},\;{\rm and}\;\; \delta=b^{23} - v^1\ee_{123}$.
 Under the action of $R$, a CUS $\KKKK$ transforms as
$$R\KKKK = R(\psi f_+) = k^1(Rf_+) + k^2(R\ee_1 f_+),$$
\noindent where we denote:
\beq Rf_+ &=& (\ap + \be\ee_{123})f_+ + (\g + \delta\ee_{123})(\ee_1 f_+),\nonumber\\
 R\ee_1f_+ &=& (\ap - \be\ee_{123})f_+ + (-\g + \delta\ee_{123})(\ee_1 f_+).
\eeq 

A matrix representation  $\rho:\clt\ri{\mt M}(2,\CC)$ of $R$ is given by: 
\beq
 \rho(R) = \left(\begin{array}{cc}
	\ap + \be i & -\g + \delta i\\
	\g + \delta i & \ap - \be i
	\end{array}\right). 
\eeq
It follows that  
\bege{\rm det}\;\rho(R) = \ap^2 + \be^2 + \g^2 + \delta^2.\enge\noi 
Under the automorphism and antiautomorphisms of $\cl_{3,0}$, respectively the graded involution, the reversion  and the conjugation,
the  multivector $R\in\cl_{3,0}$ transforms as:
\beq
\hat{R} &=& \ol{\ap} + \ol\be\ee_{12} + \ol\g\ee_{13} + \ol{\delta}\ee_{23}, \nonumber\\
\tilde{R} &=& \ol{\ap} - \ol\be\ee_{12} - \ol\g\ee_{13} - \ol{\delta}\ee_{23}, \nonumber\\
{\ul{R}} &=& {\ap} - \be\ee_{12} - \g\ee_{13} - {\delta}\ee_{23}.\eeq

\noindent Then we have the relation  
\bege R\ul{R} = \ap^2 + \be^2 + \g^2 + \delta^2 = {\rm det}\;\rho(R) \enge \noindent 
Given  $R\in\${\rm pin}_+(1,3)$, i.e., $R{\ul{R}} = 1$, we see that  det $\rho(R) = 1$ and
 $R\in {\rm SL}(2,\CC)$. Then the isomorphism
\bege\begin{array}{|l|}\hline{\rule[-3mm]{0mm}{8mm}  
\${\rm pin}_+(1,3)\simeq {\rm SL}(2,\CC)}\\ \hline\ear\enge  
\noi is explicitly exhibited.

From the condition  $R{\ul{R}} = 1$ it follows that  $\ul{R} = R^{-1}$, and 
the following transformation rules are verified in this formalism:
 \beq\label{xcv}\begin{array}{lll}
	\KKKK & \longmapsto & R \KKKK,\\
	\KKKK^* & \longmapsto & \ee_1(\ul{R\KKKK}) = \ee_1 \ul\KKKK \ul{R} = \KKKK^* R^{-1},\\
	\ol{\KKKK} & \longmapsto & \ee_1\widetilde{(R\KKKK)} = \ee_1 \tilde\KKKK \tilde{R} = \ol{\KKKK} (\ul{\hat{R}}) = \ol{\KKKK} (\hat{R})^{-1},\\
	{\ol{\KKKK}}^* & \longmapsto & (\widehat{R\KKKK}) =  \hat{R}{\ol{\KKKK}}^*,
	\end{array} 
\eeq\noi which permits to represent
\bege\label{4trw}\begin{array}{|lll|}\hline{\rule[-3mm]{0mm}{8mm}
	\KKKK } &\longmapsto&  R \KKKK\\
	\KKKK^* & \longmapsto &  \KKKK^* R^{-1}\\
	\ol{\KKKK} & \longmapsto & \ol{\KKKK} (\hat{R})^{-1}\\
	{\ol{\KKKK}}^* & \longmapsto&   \hat{R}{\ol{\KKKK}}^*\\ \hline
	\end{array} 
\enge
 \noi In this way it is proved that the transformations of CUS, CVUS, CDS and CVDS 
under $\${\rm pin}_+(1,3)\simeq {\rm SL}(2,\CC)$ are the same as pointed in the classical formalism \cite{inf,pe2}. Indeed, we see that 
\bege
\rho({\hat{R}}) = [{\rho(R)}^\dagger)]^{-1},
\enge
\noi from where expressions (\ref{4trw}) follow. 
Therefore we have the correspondence:
\beq 
\begin{array}{ccccccccc}
\KKKK & \Ll & \left(\begin{array}{cc}
		k^1 & 0\\
		k^2 & 0 \end{array}\right),& \quad & \ol{\KKKK}& \Ll & \left(\begin{array}{cc}

										0 &
 0 \\
											\ol{k}^{1'} & \ol{k}^{2'}  \end{array}\right) = 																				\left(\begin{array}{cc}
															0 & 0 \\
																	-\ol{k}_{2'} & \ol{k}_{1'}  \end{array}\right),\\
																	
	\KKKK^* & \Ll & \left(\begin{array}{cc}
		-k^2 & k^1\\
		0 & 0 \end{array}\right),& \quad & \ol{\KKKK}^*& \Ll & \left(\begin{array}{cc}
																	0 & \ol{k}_{1'} \\
																	0 & \ol{k}_{2'}  \end{array}\right).\\

																	\end{array}
																	\eeq

In order to write the four Weyl spinors (and subsequently the Dirac spinor), it is enough to consider the
 ideal $\clt f_+$ and,  using the right and left Clifford product by $\ee_1$, the reversion and the graded involution, 
the other ideals $f_+\clt$, $\clt f_-$, $f_-\clt$ are immediately obtained. The other three types of Weyl spinors are respectively 
elements of these ideals.

\section{Dirac theory in the paravector model of $\cl_{3,0}$}
\label{sec3}
In this section we introduce, according to Hestenes \cite{hes1,hes2,hes3,hes6} and Lounesto \cite{15,41},
 the algebraic description of the Dirac spinor and present the 
Dirac-Hestenes equation (DHE). 
We first reproduce some important results that are in, e.g, \cite{15,41}. 

The Dirac equation for a quantum relativistic particle of mass $m$, described by $\psi$, in a background 
with electromagnetic potential $A$,  is written as\footnote{It will be used natural units, such that $\hslash = 1$ and $c = 1$.} \cite{itz}
\bege\label{dirac}\begin{array}{|l|}\hline{\rule[-3mm]{0mm}{8mm}
\g^\mu (i\partial_\mu - e A_\mu)\psi  = (i\eth - eA) = m\psi,\quad \psi\in \CC^4}\\ \hline\end{array} 
\enge
\noi 

An element  $\Psi\in\cl_{1,3}^+\simeq {\mt M}(4,\CC)$ is written as
\bege
 {\Psi} = c +  c^{01}\g_{01} + c^{02}\g_{02} + c^{03}\g_{03} + c^{12}\g_{12} + c^{13}\g_{13} + c^{23}\g_{23}  + c^{0123}\g_{0123}.
\enge
\noi From the standard representation of $\g_\mu$ \cite{itz,gre} we have the correspondence:
\beq \rho(\Psi) &=& 
\left(\begin{array}{cccc} 
       c - ic^{12}&c^{13} - ic^{23}&-c^{03} + i^{0123}&-c^{01} + i c^{02}\\
       -c^{13} - ic^{23}&c + ic^{12}&-c^{01} - i c^{02}&c^{03} + i^{0123}\\
       -c^{03} + i^{0123}&-c^{01} + i c^{02}&c - ic^{12}&c^{13} - ic^{23}\\
       -c^{01} - i c^{02}&c^{03} + i^{0123}&-c^{13} - ic^{23}&c + ic^{12}
\end{array}\right)\nonumber\\ &=& \left(\begin{array}{cccc}
                      \phi_1 &-\phi_2^*&\phi_3 &\phi_4^*\\
                      \phi_2 &\phi_1^*&\phi_4 &-\phi_3^*\\
                      \phi_3 &\phi_4^*&\phi_1 &-\phi_2^*\\
                      \phi_4 &-\phi_3^*&\phi_2 &\phi_1^*
                      \end{array}\right) \in {\mt M}(4,\CC)
\eeq
A {Dirac spinor} $\psi$ can be expressed as an element of the left ideal  $(\CC\otimes\cl_{1,3})f$, where
 $f = \frac{1}{4}(1+\g_0)(1 + i\g_{12})$. Since we have the isomorphism
\bege
(\CC\otimes\cl_{1,3})f \simeq \clt \simeq \clt f_+ \op \clt f_-,
\enge it is easy to see that the definition of a
Dirac spinor as the sum of a CUS (an element of $\clt f_+$) and a CVDS (an element of $\clt f_-$) immediately follows.

 A Dirac spinor can also be written as
\bege
\psi = \Phi\frac{1}{2}(1 + i\g_{12})\in (\CC\otimes\cl_{1,3})f,
\enge
\noi where $\Phi = \Phi\frac{1}{2}(1 + \g_0) \in \cl_{1,3}(1+\g_0)$ 
is two times the real part of $\psi$. Therefore the following expression is obtained \cite{15,41}:
\bege
(\CC\otimes\cl_{1,3})f\ni\psi \simeq \CC\otimes\left(\begin{array}{cccc}
                      \phi_1 &0&0 &0\\
                      \phi_2 &0&0 &0\\
                      \phi_3 &0&0 &0\\
                      \phi_4 &0&0 &0
                      \end{array}\right) \simeq \CC\otimes\left(\begin{array}{c}
                      \phi_1\\
                      \phi_2\\
                      \phi_3\\
                      \phi_4
                      \end{array}\right) = \left(\begin{array}{c}
                      \psi_1\\
                      \psi_2\\
                      \psi_3\\
                      \psi_4
                      \end{array}\right)\in \CC^4,
\enge
\noi Besides, $4\;{\rm Re} (i\Phi) = \Psi\g_2\g_1$, and the
 spinor $\Phi  \in \cl_{1,3}(1+\g_0)$ is decomposed in even and odd parts:
\bege
\Phi = \Phi_0 + \Phi_1 = (\Phi_0 + \Phi_1)\frac{1}{2}(1 + \g_0) = \frac{1}{2} (\Phi_0 + \Phi_1\g_0) + \frac{1}{2} (\Phi_1 + \Phi_0\g_0),
\enge
\noi It follows that $\Phi_0 = \Phi_1\g_0$ and $\Phi_1 =  \Phi_0\g_0$. 
Taking the real part in $\CC\otimes\cl_{1,3}$ of eq.(\ref{dirac}) one obtains
\bege
\eth\Phi\g_2\g_1 - eA\Phi = m\Phi,
\enge
\noi which can be decomposed again in even and odd parts, respectively:
\beq
\eth\Phi_0\g_{21} - eA\Phi_0 &=& m\Phi_1,\nonumber\\
 \eth\Phi_1\g_{21} - eA\Phi_1 &=& m\Phi_0.
 \eeq
 \noi The {\it Dirac-Hestenes equation} is written, denoting $\Psi = \Phi_1 = \Phi_0\g_0$, as: 
 \bege\label{dihe} \begin{array}{|l|}\hline{\rule[-3mm]{0mm}{8mm}  
 \eth\Psi\g_{21} - eA\Psi = m\Psi\g_0},\quad \Psi\in \cle^+\\ \hline\end{array} 
 \enge
\noi All elements of the equation above are now multivectors of $\cl_{1,3}^+$. 

In order to simplify the notation we write
\beq
 \Psi &=& c +  c^{1}\g_{01} + c^{2}\g_{02} + c^{3}\g_{03} + b^{1}\g_{23} + b^{2}\g_{31} + b^{3}\g_{12}  + b^{0}\g_{0123}, \quad(c, b, c^\mu, b^\mu \in \RR)\nonumber\\
&=&c^0 + c^k\g_{k0} - b^k {\mathfrak{I}}\ee_k + b^0{\mathfrak{I}},
\eeq
\noi where ${\mathfrak{I}} = \ee_{123} = \g_{0123}$. 
In this way, given the electromagnetic potential $A = A^\mu\g_\mu$, 
it is valid the expression $\g_0A = A^0 + A^k\g_{0k} = A^0 - A^k\ee_k$. Denoting ${\bf{A}} = A^k\ee_k$, it follows that 
\bege\label{paravpota}
A = A^0 - {\bf{A}}.
\enge
Left multiplying eq.(\ref{dihe}) by $\g_0$,  
\bege
 \g_0\eth\Psi\g_{21} - e\g_0A\Psi = m\g_0\Psi\g_0,
 \enge
\noi and using the above notation we obtain
\bege\label{eo}
(\partial^0 - \partial^k\ee_k)\Psi{\mathfrak{I}}\ee_3 - e(A^0 - {\bf{A}})\Psi = m\g_0\Psi\g_0, \quad\Psi \in \cl_{3,0} \simeq \cl_{1,3}^+.
\enge
\noi But $\g_0\Psi\g_0$ is the {\it parity operator}, 
which will be denoted by $\Psi^P = \g_0\Psi\g_0 = {\hat{\Psi}}$. Therefore,
\bege
\hat\Psi =  c^0 - c^k\g_{k0} + b^k {\mathfrak{I}}\ee_k - b^0{\mathfrak{I}},
\enge
\noi Denoting $\partial^0 = \partial_t$ and $\phi = A^0$,   eq.(\ref{eo}) is rewritten as
\bege\label{eor}
(\partial_t - \partial^k\ee_k)\Psi{\mathfrak{I}}\ee_3 - e(\phi - {\bf{A}})\Psi = m{\hat{\Psi}}, \quad\Psi \in \cl_{3,0}, 
\enge
\noi and then
\bege\label{dicle}
\boxed{\partial_t\Psi + \nabla\Psi = \left[e({\bf{A}} - \phi)\Psi - m{\hat{\Psi}}\right] \mathfrak{I}\ee_3, \quad \Psi \in \cl_{3,0}.}
\enge
Eq.(\ref{dicle}) is the  {\bf Dirac equation} written in $\cl_{3,0}$. 
It does not contradict the impossibility of using $2\times 2$ matrices in order to describe the relativistic electron, 
because in the non-relativistic theory the wave function is represented by a complex 2-dimensional vector, with four 
real parameters, and the matrices act only by left multiplication. 
In the formalism presented in this section, $\psi$ is represented by a $2\times 2$ complex matrix, with eight real parameters,
 where the matrices act by left {\bf and/or} right multiplication.    

The element $\Psi\in\clt$ can be interpreted as the composition of three operations:  dilation, duality and Lorentz
transformations (more specifically $R$ is not a Lorentz transformation, but does {\it generate} a Lorentz transformation). 
Indeed, denoting $\g_5 = \g_{0123}$, for $\Psi\in\clt$, $\Psi{\bar{\Psi}}$ is expressed as $a + \g_5b = \rho e^{\g_5\be},\;\rho > 0$, and we can express \cite{hes1,hes2,hes3}
\bege\label{dh}
\Psi = {\sqrt{\rho}}\; e^{\g_5\be/2} R,\enge\noi where $R{\bar R} = 1$ and $\be$ is the Takabayasi angle \cite{yv,tak}. Consequently, 
$R\in \${\rm pin}_+(1,3)$ and $R$ generates a Lorentz transformation \cite{15}. 

\subsection{The plane wave solution of the Dirac equation in $\clt$}

The Dirac equation (eq.(\ref{dicle})) is written, in the absence of external fields, as:  
\bege\label{ploiuy}
\partial_t\Psi + \nabla\Psi = - m{\hat{\Psi}} {\mathfrak{I}}\ee_3, \quad \Psi \in \cl_{3,0}.
\enge 
Since  eq.(\ref{ploiuy}) is invariant under Lorentz transformations, 
we first solve the equation in a rest referential and, after this, a {\it boost} $L$ is applied. 
In the Pauli algebra $\clt$, the action of the momentum operator ${\mathfrak{p}}$ on the wave function $\Psi\in\clt$ is given by
\bege
{\mathfrak{p}}\Psi = \nabla\Psi\ip\ee_3.
\enge\noi The eigenvector of ${\mathfrak{p}}$ is ${\bf p}\in\RR^3$ and in the rest referential, 
${\bf p} = 0$. Then  eq.(\ref{ploiuy}) is written as
\bege\label{sem}
\partial_t\Psi = - m{\hat{\Psi}} {\mathfrak{I}}\ee_3
\enge \noi Consider the solution of eq.(\ref{sem}) to be of the type
\bege
\Psi = \Psi_0\;{\rm exp}(-{\mathfrak{I}}\ee_3\om).
\enge\noi Substituting this solution is eq.(\ref{sem}) we obtain
\bege
\Psi_0\om = m {\hat{\Psi}_0}.
\enge\noi It follows that, in the case of even multivectors, the relation $\Psi_0 = {\hat{\Psi}}_0$ is satisfied, and in 
these conditions $\om = m$. For odd multivectors, $\Psi_0 = - {\hat{\Psi}}_0$ and consequently  $\om = -m$. 

We first investigate elements of  $\Lambda^0(\RR^3)\op\Lambda^2(\RR^3)\hko\clt$.  
For $\Psi_0 \in \Lambda^0(\RR^3) $, the solution is given (up to scalars) by:
\bege\label{werta}
\Psi = {\rm exp}(-\ip\ee_3 mt).
\enge\noi On the other hand, for
 $\Psi_0 \in \Lambda^2(\RR^3)$, there are three possibilities:
 $\Psi_0 = \ee_1\ee_2$, $\Psi_0 = \ee_1\ee_3$ and $\Psi_0 = \ee_2\ee_3$ (the general case is obtained by linearity).
\begin{enumerate}
\item The expression $\ee_1\ee_2  = \ip\ee_3 = {\rm exp}(\ip\ee_3\pi/2)$ shows that the choice $\ee_1\ee_2$ only adds the phase factor 
to the spinor (wave function).
\item The choice $\Psi_0 = \ee_2\ee_3$ is obtained from  $- (\ee_1\ee_2) (\ee_1\ee_3) = \ee_2\ee_3$, and it 
only adds a phase factor to the third choice:
\item $\Psi_0 = \ee_1\ee_3$.  
\end{enumerate} Then the second solution of eq.(\ref{sem}) is given by 
 \bege\label{puo}
\Psi = \ee_1\ee_3\;{\rm exp}(-\ip\ee_3 mt).
\enge\noi The two solutions given by eqs.(\ref{werta}) and (\ref{puo}) are positive frequency solutions \cite{dirac,itz}. 

In order to obtain the other two solutions of the Dirac equation in $\clt$, we must investigate elements of 
$\Lambda^1(\RR^3)\op\Lambda^3(\RR^3)\hko\clt$. For $\Psi_0 \in \Lambda^3(\RR^3)$, we have the solution:
\bege\label{best}
\Psi = \ip\;{\rm exp}(\ip\ee_3 mt).
\enge\noi For the fourth solution, let  $\Psi_0 \in \Lambda^1(\RR^3)$. Then the solution is obtained if eq.(\ref{best}) 
is left multiplied by $\ee_1\ee_2$, $\ee_2\ee_3$ or $\ee_1\ee_3$, since the general case is obtained by linearity. 
From the same reason cited in the last paragraph, after eq.(\ref{werta}), all the choices other than $\ee_1\ee_3$ are
 redundant, in the sense that they only add a phase factor in $\Psi$. 
With the choice $\ee_1\ee_3$, we have  $\ee_1\ee_3 \ip = \ee_2$. Then the fourth solution of 
eq.(\ref{sem}) is given by 
\bege
\Psi = \ee_2\;{\rm exp}(\ip\ee_3 mt).
\enge
A general solution of eq.(\ref{sem}) is given by the linear span of the four solutions obtained, listed below:
\bege \label{solp}\bea{|l|}\hline{\rule[-3mm]{0mm}{8mm}\bea{l}
 \Psi^{(+\uparrow)} = {\rm exp}(-\ip\ee_3 mt)\\
 \Psi^{(+\downarrow)} =  \ee_1\ee_3\;{\rm exp}(-\ip\ee_3 mt)\\
 \Psi^{(-\uparrow)} = \ip\;{\rm exp}(\ip\ee_3 mt)\\
 \Psi^{(-\downarrow)} =  \ee_2\;{\rm exp}(\ip\ee_3 mt) \ear}\\ \hline\ear
\enge
It is worth to note that the general solution of  $\Psi$
 is given, up the phase factor $\Psi \mapsto \Psi{\rm exp}(\ap\ip\ee_3)$. 
The general solution is given by the linear combination (with $\CC$-coefficients) of the solutions
given by eqs.(\ref{solp}). The $\CC$-coeficients are of the form
\bege
(c + d\ip\ee_3),\qquad c, d\in \RR.
\enge\noi They {\bf right} multiply the functions. The {\bf left}  linear combination {\bf is forbidden}, 
since the operator $\nabla$ does not commute with this possibility. 
For a particle with momentum  ${\bf{p}}$ we obtain the solution if the boost $L = L({\bf{p}})$
is applied \cite{las}:
\bege \bea{|l|}\hline{\rule[-3mm]{0mm}{8mm}\bea{l}
 \Psi^{(+\uparrow)} = L({\bf{p}})\;{\rm exp}[-\ip\ee_3 (Et - {\bf{p}}\cdot {\bf{x}})]\\
 \Psi^{(+\downarrow)} = L({\bf{p}})\; \ee_1\ee_3\;{\rm exp}[-\ip\ee_3 (Et - {\bf{p}}\cdot {\bf{x}})]\\
 \Psi^{(-\uparrow)} = L({\bf{p}})\;\ip\;{\rm exp}[\ip\ee_3 (Et - {\bf{p}}\cdot {\bf{x}})]\\
 \Psi^{(-\downarrow)} =  L({\bf{p}})\;\ee_2\;{\rm exp}[\ip\ee_3 (Et - {\bf{p}}\cdot {\bf{x}})] \ear}\\ \hline\ear
\enge
\subsection{Dirac spinors}
Penrose denotes a Dirac spinor as an element of $\GG^A\oplus\GG_{A'}$ \cite{pe2}. 
Since the Dirac spinor has four $\CC$-components,
 it suggests that the {Dirac spinor can be described as a multivector of $\clt$}.
 In the present formalism, the Dirac spinor  $\psi$ is an element of $\clt f_+ \oplus \clt f_- \simeq \clt$. Indeed,
\beq
\clt\ni\psi &=& \psi (f_+ + f_-) = \psi f_+ + \psi \ee_1 f_+\ee_1\nonumber\\
             &=& \KKKK + \LL\ee_1,
             \eeq
             \noi where
             \bege\label{tesd}
              \KKKK = \psi f_+\;\; {\rm and}\;\; \LL = \psi (\ee_1 f_+).
              \enge

 \subsection{Decomposition of the Dirac equation in terms of Weyl spinors}

The  Dirac equation in $\clt$, eq.(\ref{dicle}), is led to two Weyl equations \cite{gre}.  
Indeed, consider  eq.(\ref{dicle}) without external fields:
\bege\label{dic1}
(\pa_t + \nabla)\psi\ip\ee_3 = m{\hat{\psi}}, \quad \psi\in\clt.
\enge
\noi Besides, consider the decomposition
 $\psi = \psi f_+ + \psi f_-$, where $f_\pm = \frac{1}{2}(1 \pm \ee_3)$. We write
\beq
\xi &:=& \psi f_+ \in \clt f_+\nonumber\\
{\hat{\eta}} &:=& \psi f_- \in \clt f_-
\eeq
\noi since ${\widehat{f_+}} = f_-$. 
The correspondence with the  notation used in the last subsection is given by 
\bege\xi = \KKKK\;\;{\rm  and}\;\; {\hat{\eta}} = \LL\ee_1.\enge\noi It follows that
\bege
(\pa_t + \nabla)(\xi + {\hat{\eta}})\ip\ee_3 = m({\hat{\xi}} + \eta),
\enge\noi and then
\bege
(\pa_t + \nabla)\xi\ip - (\pa_t + \nabla) {\hat{\eta}}\ip = m{\hat{\xi}} + m\eta,
\enge
\noi where the relations $\xi\ee_3 = \xi$ and ${\hat{\eta}}\ee_3 = - {\hat{\eta}}$ follows from the fact that
 $\xi\in \clt f_+$ and ${\hat{\eta}}\in \clt f_-$.

Separating the terms in $\clt f_+$ and in $\clt f_-$ two equations are obtained:
\beq
(\pa_t + \nabla) \xi \ip &=& m\eta,\\
-(\pa_t +\nabla) {\hat{\eta}}\ip &=& m{\hat{\xi}}.
\eeq
Taking the graded involution of the last equation, the Dirac equation in terms of the  Weyl spinors gives the following coupled system:
\bege\begin{array}{|l|}\hline{\rule[-3mm]{0mm}{8mm} 
(\pa_t + \nabla) \xi \ip = m\eta}\\
(\pa_t -\nabla) {\eta}\ip = m\xi\\ \hline\ear\enge\noi This result is presented, e.g., in \cite{itz,gre}. 
\subsection{Decomposition of the Dirac equation in terms of Pauli spinors}
A multivector  $\psi\in\clt$ is written as
\beq
\clt\ni\psi &=& a + a^i\ee_i + a^{ij}\ee_{ij} + p \ee_{123}\nonumber\\
            &=&a + a^{12}\ee_{12} + a^{23}\ee_{23} + a^{13}\ee_{13} + (a^3 + a^{1}\ee_{13} + a^2\ee_{23} + p\ee_{12})\ee_3\nonumber\\
            &=&\phi +  \chi\ee_3, \quad \phi,\chi \in \clt^+.
            \eeq
            
   \noi If we substitute in eq.(\ref{dic1}) it follows that
   
\bege
(\pa_t + \nabla)(\phi + \chi\ee_3)\ip\ee_3 = m(\phi - \chi\ee_3),
\enge              
\noi  This equation is separated in even and odd parts:
\bege\begin{array}{|l|}\hline{\rule[-3mm]{0mm}{8mm}  
\pa_t\phi\ip\ee_3 + \nabla\chi\ip = m\phi}\\
\pa_t\chi\ip\ee_3 + \nabla\phi\ip = -m\chi\\ \hline\end{array}
\enge
\noi a system of two coupled equations.

\subsection{Paravectors of Minkowski spacetime obtained from Weyl spinors in $\cl_{3,0}$}
\label{subsec{3.3}}
An arbitrary paravector \cite{bay2,bayoo,por1} $\vv\in\RR\op\RR^3$ is written as
\bege
\vv := \KKKK\ee_1\ol{\KKKK} = \KKKK\ee_1 \ee_1\tilde{\KKKK} = \KKKK\tilde{\KKKK}.\enge\noi 
From eqs.(\ref{scta}) and (\ref{sctp}), we obtain:
  \bege\label{kkj}\KKKK\tilde{\KKKK} = 
                \left(\begin{array}{ll} k^1  \ol{k}^{1'} & k^1  \ol{k}^{2'}\\
		k^2  \ol{k}^{1'} & k^2  \ol{k}^{2'}
		 \end{array}\right).
	\enge	 																	
																	
Given \bege
 \KKKK = k^1f_+ + k^2 \ee_1 f_+\in \clt f_+\enge \noi and
\bege \ol{\KKKK} = \ol{k}^{1'}(f_-\ee_1) + \ol{k}^{2'}(f_-)\in \clt f_-,
\enge\noi we have the expression
 \bege\label{kkt}\begin{array}{|l|}\hline{\rule[-3mm]{0mm}{8mm}
 \KKKK\tilde{\KKKK} = \KKKK\ee_1\ol{\KKKK} = k^1\ol{k}^{1'}f_+ +  k^1 \ol{k}^{2'}f_+\ee_1	+	k^2  \ol{k}^{1'} f_-\ee_1 +  k^2  \ol{k}^{2'}f_-}\\ \hline\ear
\enge
\noi Using the representation  (\ref{rep11}) in eq.(\ref{kkj}), the correspondence 
\bege\begin{array}{|l|}\hline{\rule[-3mm]{0mm}{8mm}
f_+ \Ll o^Ao^{A'},\quad\quad f_+\ee_1 \Ll o^Ai^{A'},\quad\quad f_-\ee_1 \Ll i^Ao^{A'},\quad\quad f_- \longleftrightarrow i^Ai^{A'}}\\ \hline\ear
\enge\noi is obtained.
The idempotents of $\cl_{3,0}$, constructed from the vectors $\ee_1$ and  $\ee_3$, 
can be identified with the null tetrad given in \cite{pe2}. 
In this way, the null tetrad is constructed from spinors in Minkowski spacetime using the Clifford algebra $\clt$. 

A paravector $\mma \in \RR \oplus\RR^3 \in \cl_{3,0}$ can be written as  
 \bege\begin{array}{|l|}\hline{\rule[-3mm]{0mm}{8mm}\mma = 2\KKKK\ee_1\ol{\KKKK} =  2\KKKK\tilde{\KKKK}}\\ \hline\ear\enge
\noi Indeed, an operatorial spinor $\psi\in\cl_{3,0}^+$ is given by,
\bege\label{mma1}\begin{array}{lll}\mma = 2\KKKK{\tilde{\KKKK}} &=& 2\psi f_+ f_+{\tilde{\psi}} = 2\psi f_+{\tilde{\psi}} =  \psi(1 + \ee_3){\tilde{\psi}}\\{} &=& \psi{\tilde{\psi}} + \psi\ee_3{\tilde{\psi}} \\&=& \mma^0 + \mma^i\ee_i, \quad\quad\quad (i = 1, 2, 3).\ear
\enge
\noi The paravector $\mma$ points to the future: 
\beq
 \psi{\tilde{\psi}} &=& \mma^0\n
 &=& (a + b\ee_{12} + c\ee_{13} + d\ee_{23})(a - b\ee_{12} - c\ee_{13} - d\ee_{23})\n
 &=& a^2 + b^2 + c^2 + d^2 \n
&>& 0.
\eeq
\noi Besides, from the relation  $(\mma^i)^2 = (\psi{\bar{\psi}})^2 = (\mma^0)^2$, the paravector $\mma$ is null. Indeed,
\bege
\mma^2 := (\mma^0)^2 - (\mma^i)^2 = 0.
\enge\noi   
The last expression in eq.(\ref{mma1}) follows from the property that it always possible to write ${\bf x}\in\RR^3$ as: 
\bege\label{puyt}
{\bf{x}} = x^i\ee_i = \psi\ee_3{\tilde{\psi}},
\enge
\noi which is the composition of a rotation with a dilation \cite{hes3}. 
Eq.(\ref{puyt}), multiplied by  $\hbar/2$, defines the spin density.

From eq.(\ref{kkt}), two paravectors $\mma$ and $\mmb$ are written as:
\beq
 \mma &=& k^1\ol{k}^{1'}f_+ +  k^1 \ol{k}^{2'}f_+\ee_1	+	k^2  \ol{k}^{1'} f_-\ee_1 +  k^2  \ol{k}^{2'}f_-\nonumber\\
 &=& \mma_0 + \mma^i\ee_i,\nonumber\\
\mmb &=&  r^1\ol{r}^{1'}f_+ +  r^1 \ol{r}^{2'}f_+\ee_1	+	r^2  \ol{r}^{1'} f_-\ee_1 +  r^2  \ol{r}^{2'}f_-\nonumber\\
 &=& \mmb_0 + \mmb^i\ee_i,\eeq
 and their respective conjugation are given by
 \beq
 \hat\mma &=& {\widehat{k^1{\ol{k}}^{1'}}}f_- -  {\widehat{k^1 {\ol{k}}^{2'}}}\ee_1f_-	-	{\widehat {k^2  {\ol{k}}^{1'}}} \ee_1f_+ +  {\widehat{k^2 
 {\ol{k}}^{2'}}}f_+\nonumber\\ &=& \mma_0 - \mma^i\ee_i,\nonumber\\
\hat\mmb &=& {\widehat{r^1\ol{r}^{1'}}}f_- -  {\widehat{r^1 \ol{r}^{2'}}}\ee_1f_-	-
	{\widehat {r^2  \ol{r}^{1'}}} \ee_1f_+ +  {\widehat{r^2  \ol{r}^{2'}}}f_+\nonumber\\
 &=& \mmb_0 - \mmb^i\ee_i.
\eeq
The Clifford relation, for paravectors, is naturally obtained:
\beq
\mma{\hat{\mmb}} + \mmb{\hat{\mma}} &=&  (k^1\ol{k}^{1'}r^2  \ol{r}^{2'} +  k^1 \ol{k}^{2'}r^2  \ol{r}^{1'}	+	k^2  \ol{k}^{1'}r^1 \ol{r}^{2'} +  k^2  \ol{k}^{2'}r^1\ol{r}^{1'}) \nonumber\\
 &=& 2(\mma^0 \mmb^0 - \mma^i\mmb_i)\n
 &=& 2g(\mma, \mmb).
\eeq

\section{The Clifford algebra $\cl_{0,3}\simeq\HH\oplus\HH$}

Consider  $\RR^{0,3}$ and an orthonormal frame field $\{\mme_1, \mme_2, \mme_3\}$.  The Clifford algebra 
$\cl_{0,3}$ is generated by $\{1, \mme_1, \mme_2, \mme_3\}$, that satisfies 
\bege
g(\mme_i, \mme_j) = -\delta_{ij} = \me(\mme_i\mme_j + \mme_j\mme_i), \quad (i,j = 1, 2, 3).
\enge\noi In particular $\mme^2 = -1$. 

We first take the redundant dimensions out of the formalism, proving the 
\medbreak{\bf Proposition} $\vvn$ $\cl_{0,3}\mmf_+ \simeq \cl_{0,3}^+\mmf_+$, where $\mmf_\pm = \me(1 \pm \ip)$ and $\ip:=\mme_1\mme_2\mme_3$. $\vbn$\medbreak 
\noi{\bf Proof}: The left minimal ideal  $\cl_{0,3}\mmf_+$ is isomorphic to $\HH$, as an algebra. Besides, an arbitrary element of  $\cl_{0,3}$ is written as
\beq
 A &=& a^0 + a^k\mme_k + b^1\mme_{23} + b^2\mme_{31} + b^3\mme_{12} + b^0\mme_{123}\nonumber\\ 
&=& a^0 + a^k\mme_k - b^k\mme_k\mme_{123} + b^0\mme_{123}.
\eeq \noi It is easily seen that
\beq
A\mmf_+ &=& [(a^0 + b^0) + (a^k - b^k)\mme_k]\mmf_+\nonumber\\
&=& [(a^0 + b^0) + (a^k - b^k)\mme_k\mme_{123}]\mmf_+
\eeq\noi Therefore, given $A\mmf_+\in\cl_{0,3}\mmf_+$, and writing
 $A' = (a^0 + b^0) + (a^k - b^k)\mme_k\mme_{123}$, we see that 
 $A'\in\cl_{0,3}^+$ and that $A\mmf_+ = A'\mmf_+$. 
This shows that  $\cl_{0,3}\mmf_+ \hko \cl_{0,3}^+\mmf_+$. The another inclusion follows immediately, since $\cl_{0,3}^+$ is the even subalgebra of 
$\cl_{0,3}$. \bfr$\Box$\efr

Now consider an even element $Q\in\cl_{0,3}^+$ given by 
\beq\label{q}  Q &=& a + b\mme_{12} + c\mme_{13} + d\mme_{23}\nonumber\\
                      &=&(a + b\mme_{12}) + \mme_{13}(c - d\mme_{12}) \nonumber\\
                      &=& k^1 + \mme_{13} k^2.\eeq 

\noi Another  possibility to describe Weyl spinors is to consider the algebra  $\cl_{0,3} \simeq \HH \oplus \HH$. 
A spinor $\KKKK = Q\mmf_+\in \cl_{0,3} \mmf_+$ is expressed as a 
\medbreak$\bullet$ {\ul{\it Contravariant undotted spinor}} (CUS):
\bege\bea{|l|}\hline{\rule[-3mm]{0mm}{8mm}\KKKK = (k^1\mmf_+ + \mme_{13}k^2\mmf_+)}\\ \hline\ear
\enge \noi 
We also define the
\medbreak$\bullet$ {\ul{\it Contravariant dotted spinor}} (CDS):
\bege\bea{|l|}\hline{\rule[-3mm]{0mm}{8mm}{\bar\KKKK} = (\mmf_+{\bar k^1} - \mmf_+{\bar k^2}\mme_{13})}\\ \hline\ear
\enge

 \noi Left multiplying the conjugate of  $\KKKK$ by $\mme_{13}$, we obtain
\beq \mme_{13}\ul{\KKKK} &=& \mme_{13}(\mmf_+\ol{k^1} + \mmf_+\ol{k^2}\mme_{31})\n
 &=& \mmf_+ k^1 \mme_{13} + \mmf_+ k^2.\eeq\noi When the last expression is multiplied by another spinor $\eta\in\cl_{0,3}f_+$, giving 
\bege\mme_{13}{\ul\KKKK}\eta = k^1 \ol{\eta^1} \mme_{13}\mmf_+ - k^1\eta^2\mmf_+ + k^2\eta^1\mmf_+ + k^2\ol{\eta^2}\mme_{13}\mmf_+,\enge\noi the spinor metric is
obtained in $\cl_{0,3}$:
\bege\bea{|l|}\hline{\rule[-3mm]{0mm}{8mm}
{\mt G}(\KKKK,\eta) := 2\langle(\ee_{13}\ul{\KKKK})\eta\rangle_0 = (k^2\eta^1 - k^1\eta^2)\mmf_+}\\ \hline\ear\enge

Now consider the application $\si:\cl_{0,3}^+\ri\cl_{0,3}^+$ given by
\bege\si(Q) = \mme_{32}\ul{Q}\mme_{23},\enge where $\ul{Q} = \widehat{\tilde{Q}}.$ 
 The map $\si$ takes $\cl_{0,3}^+$, viewed as a left-module, onto $\cl_{0,3}^+$, but now viewed as a right-module. Indeed, from eq.(\ref{q}) we have:
\beq \si(k^1 + \mme_{13}k^2) &=& \si(a + b\mme_{12} + c\mme_{13} + d\mme_{23})\nonumber\\
&=& \mme_{32}(a + b\mme_{12} + c\mme_{13} + d\mme_{23})\mme_{23}\nonumber\\
&=&(a + b\mme_{12}) + (c - d\mme_{12})\mme_{13}\nonumber\\
&=&k^1 + k^2\mme_{13}.
\eeq
For a spinor $\KKKK = Q\mmf_+\in\cl_{0,3}^+\mmf_+$, it follows that
\beq\si(\KKKK) = \si (Q \mmf_+) &=& \mme_{32}(\ul{Q\mmf_+})\mme_{23}\nonumber\\
 &=& \mme_{32}(\mmf_+\ul{Q}\mme_{23})\nonumber\\
&=& \mmf_+\mme_{32}\ul{Q}\mme_{23}\nonumber\\ &=& \mmf_+ \si(Q)\nonumber\\
&=& \mmf_+ (k^1 + k^2 \mme_{13}).\eeq
\noi In this way,
\beq\si(\KKKK)\mme_{13} &=& \mmf_+ (k^1 \mme_{13} - k^2)\nonumber\\ 
&=& \mmf_+ (-k^2 + k^1\mme_{13})\nonumber\\
 &=& \KKKK^*.\eeq

 The {spinor metric} is alternatively defined as:
\bege \bea{|l|}\hline{\rule[-3mm]{0mm}{8mm}{\mt G}(\psi, \phi) := \langle\si(\psi)\mme_{13}\phi\rangle_{0\op 3} = \frac{1}{2}(
\si(\psi)\mme_{13}\phi + \mme_{21} \si(\psi)\mme_{13} \phi \mme_{12})}\\ \hline\ear\enge
\noi The algebra $\cl_{0,3}$ is not so natural as $\cl_{3,0}$ to describe a  lorentzian spacetime. 
Indeed it is suitable to investigate an euclidian space $\RR^4$, since given $\uu\in\RR^4$ we have,
\bege
\uu{\bar{\uu}} = u_0^2 + {\vec u}^2, \quad u_0\in\RR,\;{\vec u}\in\RR^3.
\enge
\noi Besides, $\cl_{0,3}\simeq \HH\op\HH$ is a semi-simple algebra, and the ring $\HH$ is not commutative. It is then necessary to 
treat the right and left product by $\HH$. We proved that there exists an application $\si:\cl_{0,3}^+\rightarrow\cl_{0,3}^+$
 that maps a left $\HH$-module onto a right $\HH$-module.  

\section{Concluding Remarks}We introduced the covariant and contravariant, dotted and undotted Weyl spinors in $\clt$ and the two last ones in $\cl_{0,3}$,
where we constructed an application that  maps $\cl_{0,3}^+$, viewed as a left $\HH$-module, onto $\cl_{0,3}^+$, but now viewed as a right $\HH$-module.
The correspondence between the idempotents that generate the four lateral minimal ideals in $\clt$ is obtained, if the antiautomorphisms in $\clt$
act on the four types of Weyl spinors, respctively elements of the minimal lateral ideals in $\clt$. The null tetrad is 
obtained in the paravector model of Minkowski spacetime.
The Dirac equation in $\clt$ is also presented and discussed. The plane wave solutions of such equation are constructed and the Dirac theory is formulated 
using the $\clt$ structure, where operators, vectors and tensors are
unified described by multivectors and multiforms in the Clifford formalism. The decomposition of the Dirac equation into a system of coupled equations,
written in terms of Weyl, and alternatively, Pauli spinors, is also presented.

\section*{Acknowledgements}
The authors are greatly indebted to Dr. R. A. Mosna for pointing out some mistakes and for giving many other helpful
 suggestions.

\end{document}